# Semi-Dirac Semimetal in Silicene Oxide

Chengyong Zhong,[1] Yuanping Chen,[1,*] Yuee Xie,[1] Yi-Yang Sun,[2,*] Shengbai Zhang[2]

[1]*School of Physics and Optoelectronics, Xiangtan University, Xiangtan, Hunan 411105, China*

[2]*Department of Physics, Applied Physics, and Astronomy, Rensselaer Polytechnic Institute, Troy, New York 12180, USA*

[*]Emails: chenyp@xtu.edu.cn (Y. Chen), suny4@rpi.edu (Y.-Y. Sun)

## ABSTRACT

Semi-Dirac semimetal is a material exhibiting linear band dispersion in one direction and quadratic band dispersion in the orthogonal direction and, therefore, hosts massless and massive fermions at the same point in the momentum space. While a number of interesting physical properties have been predicted in semi-Dirac semimetals, it has been rare to realize such materials in condensed matters. Based on the fact that some honeycomb materials are easily oxidized or chemically absorb other atoms, here, we theoretically propose an approach of modifying their band structures by covalent addition of group-VI elements and strain engineering. We predict a silicene oxide with chemical formula of $Si_2O$ to be a candidate of semi-Dirac semimetal. Our approach is backed by the analysis and understanding of the effect of *p*-orbital frustration on the band structure of the graphene-like materials.

Dirac Semimetals, as represented by graphene, have been intensively studied in the past decade as a condensed matter platform of massless Dirac fermions [1, 2]. The electronic structure of graphene is characterized by two Dirac points, located at *K* and *K'*, respectively, in the momentum space [3]. In a tight-binding (TB) picture, by tuning the nearest-neighbor hopping energies in a graphene lattice, as illustrated in Figure 1, the two Dirac points can approach each other and merge into one forming the so-called semi-Dirac point, near which the band dispersion exhibits a peculiar feature, i.e., being linear in one direction and quadratic in the orthogonal direction [4-6]. Materials possessing semi-Dirac points are called semi-Dirac semimetals, which provide a platform where massless and massive Dirac fermions coexist. Besides the apparent highly anisotropic transport properties [7], a number of other interesting properties have been predicted for the semi-Dirac semimetals, such as distinct Landau-level spectrum in a magnetic field [8, 9], non-Fermi liquid [10], Anderson localization [11] and Bloch-Zener oscillations [12]. It is therefore of great interest to search for materials realizing the semi-Dirac semimetals.

While the TB picture provides important guidance, the search for semi-Dirac semimetals still relies on non-trivial materials design. A $VO_2$/$TiO_2$ superlattice has been theoretically proposed to hold multiple semi-Dirac points within the first Brillouin zone (BZ) [13]. The formation mechanism of the semi-Dirac points in this material is described by a TB model [14] different from that illustrated in Figure 1. Recently, black phosphorus (BP) has been predicted to exhibit a single semi-Dirac point in the BZ under pressure [15]. An exciting experiment has shown that a giant Stark effect through electron doping on the surface of BP indeed yields a semi-Dirac point [16], albeit at heavily *n*-type doped condition with the semi-Dirac point below the Fermi level by about 0.5 eV. A Dirac-to-semi-Dirac transition by merging two Dirac points, as illustrated in Figure 1, still remains to be demonstrated in a solid-state system, even though it has been observed in systems such as ultra-cold atoms trapped in a periodic potential [17] and a photonic crystal [18]. A seemingly straightforward approach to realize the semi-Dirac semimetals is to start with the honeycomb lattice and break the hexagonal symmetry, e.g., by strain, so that the hopping energies $t_2 = 2t_1$ (Figure 1). However, directly applying strain to realize the transition in materials such as graphene or silicene is prohibited by the tremendous strain requested, which will disintegrate the material [19, 20]. On the other hand, some successfully synthesized graphene-like honeycomb materials, such as silicene[21-24], germanene[25] and stanene[26], are found to be easily oxidized or chemically absorb other atoms because of their buckling geometries. Obviously, the absorbed atoms will greatly change the hopping energies in the honeycomb lattice. It is nature to ask can we use the absorbed atoms to tune the honeycomb structures from Dirac semimetals to semi-Dirac semimetals.

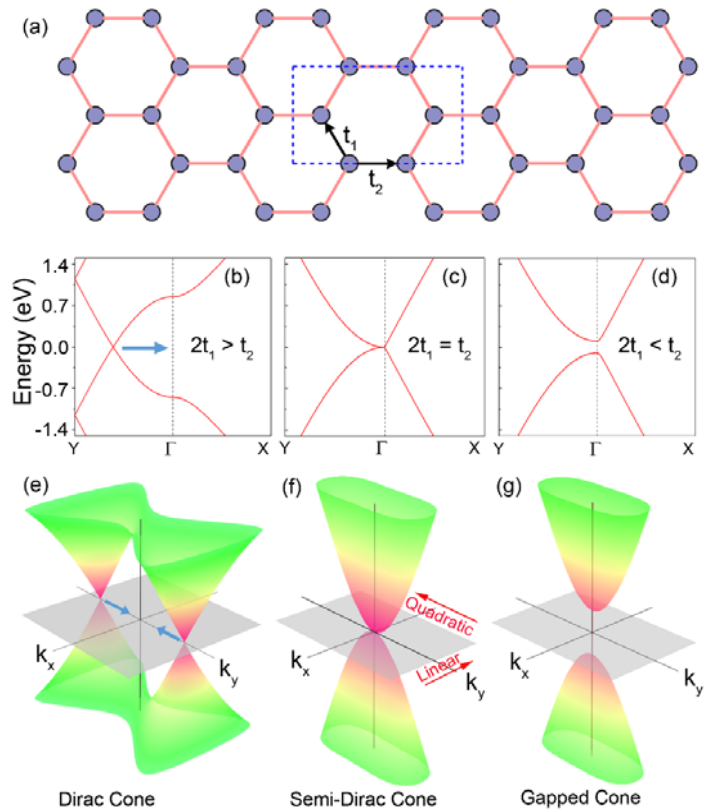

**Figure 1.** (a) A honeycomb lattice with a rectangular unit cell (indicated with blue dashed lines). Two

tight-binding (TB) hopping parameters, $t_1$ and $t_2$, are marked in the unit cell, which are equivalent in a perfect honeycomb lattice. (b)-(d) The band structures of the honeycomb lattice from the TB model under three different conditions between $t_1$ and $t_2$. (e)-(g) Three-dimensional plots of the band structures corresponding to (b)-(d), respectively. The blue arrows in (b) and (e) indicate the shift of Dirac points towards the Γ point as $t_2$ increases relative to $t_1$.

In this paper, using first-principles computation, we search for semi-Dirac semimetals through synergic tuning of the chemical and structural degrees of freedom in honeycomb lattices. The covalent addition of group-VI atoms (O, S, and Se) to silicene, germanene, and stanene are considered as examples. The results indicate that this indeed leads to the merging of two Dirac points into one highly anisotropic semi-Dirac point. In particular, covalent addition of bridging atoms is employed to enhance the strength of $t_2$. Then, structural strain is applied to fine-tune the electronic structure achieving the semi-Dirac point. The oxygen modified silicene (or silicene oxide) is found to be a promising candidate of semi-Dirac semimetal.

Our first-principles calculations were performed within the density functional theory (DFT) [27, 28] formalism as implemented in the VASP code [29, 30]. We used the generalized gradient approximation of Perdew-Burke-Ernzerhof (PBE) [31]. The interaction between the core and valence electrons was described by the projector-augmented wave method [32]. The kinetic energy cutoff of 500 eV, 350 eV and 300 eV were used for the oxides, sulfides and selenides on graphene, silicene, germanene, and stanene, respectively. The atomic positions were optimized using the conjugate gradient method with allowed maximum force of $10^{-2}$ eV/Å. A 7×11×1 $k$-point grid according to Monkhorst-Pack scheme was used to sample the Brillouin zone (BZ). The vacuum region between adjacent images in the direction normal to the silicene plane was kept at about 15 Å. When applying uniaxial strain along *a* axis, we changed the length of *a* axis while letting the *b* axis to relax fully. In the calculation of phonon spectra, we employed the finite difference method with a 3×3×1 supercell using the Phonopy code [33] with the forces calculated from VASP.

We first apply the tight-binding (TB) model of a honeycomb lattice in a rectangular unit cell, which is described by two different nearest-neighbor hopping energies, $t_1$ and $t_2$, as shown in Figure 1a. In pristine graphene or graphene-like materials, e.g., silicene, $t_1 = t_2$. There are two Dirac points located at ±2/3 Γ-Y (Figure 1b) in a rectangular BZ, which are folded from the *K* and *K'* points in a hexagonal BZ. If increasing the strength of $t_2$, the two Dirac points will approach each other until reaching the critical condition $t_2 = 2t_1$, where the two Dirac points merge into one semi-Dirac point (Figure 1c). Further increasing $t_2$ will open a gap at the Γ point (Figure 1d). According to our TB results, in order to merge the two Dirac points in graphene or silicene to a semi-Dirac point, the applied strain along the armchair direction would exceed 20%, which cannot be sustained in these materials. Thus, it is hardly realizable in graphene or silicene to form a semi-Dirac semimetal simply with the aid of strain.

It is well known that graphene can form graphene oxide, where oxygen atoms can adopt the bridge sites forming epoxide structure [34-36]. This provides an efficient chemical approach to tune the hopping energy $t_2$. We first screen such materials by adsorbing group-VI atoms (O, S, Se) on the bridge sites of graphene, silicene, germanene and stanene. We use DFT calculations with the Perdew-Burke-Ernzerhof (PBE) functional and a structure shown in Figure 2a. Our results are summarized in Table I. For the oxides, graphene oxide opens a large band gap, indicating an overly enhanced $t_2$, while germanene and stanene oxides remain to be Dirac semimetal with two Dirac points in the BZ. Only silicene oxide is close to be a semi-Dirac semimetal (Figure 2b), which will be discussed below in detail. For the sulfides and selenides, graphene cannot form such compounds based on our calculation results due to the unfavorable triangular bonding. Both sulfide and selenide of silicene open a relatively large band gap, while the sulfides and selenides of germanene and stanene are metals. The band structures of all these compounds are shown in Figure S1 in the Supplementary Information (SI).

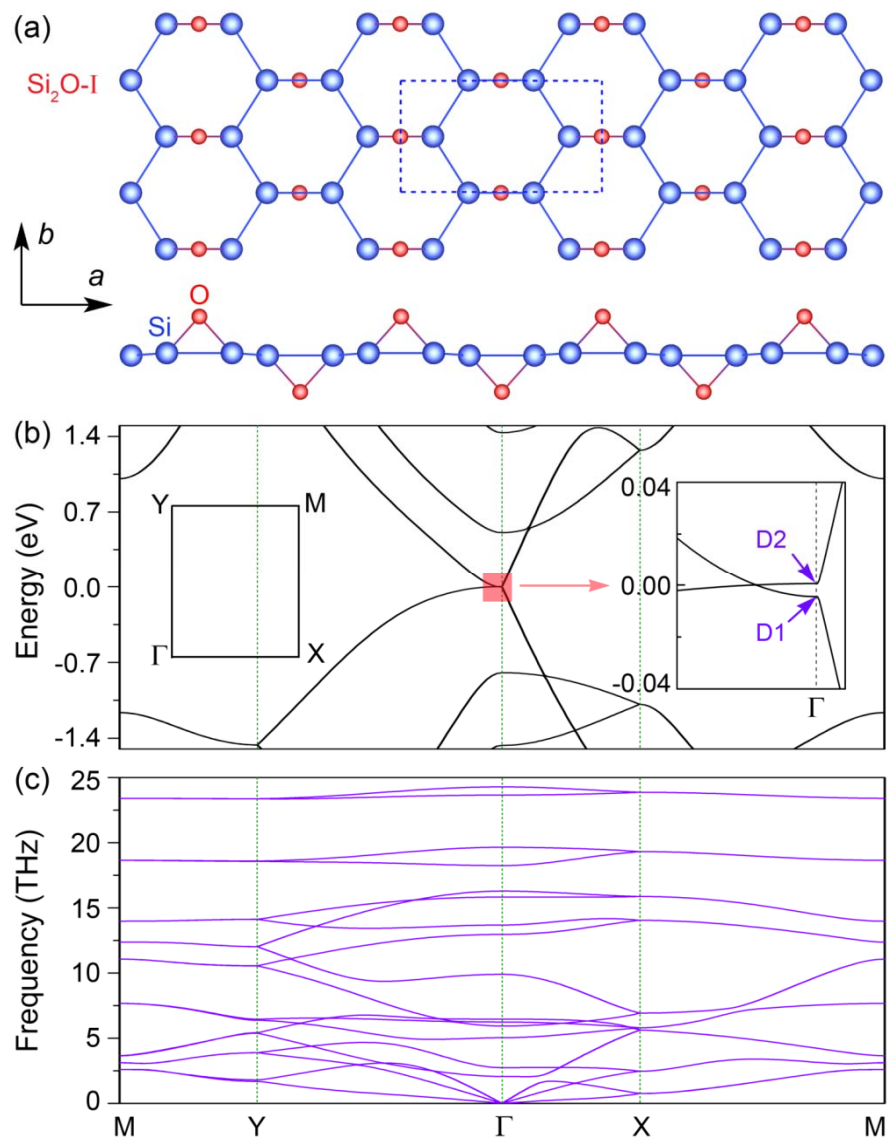

**Figure 2.** (a) Atomic structure of silicene oxide, termed as Si$_2$O-I, with top view shown at the top and side

view at the bottom. The blue and red spheres represent silicon and oxygen atoms, respectively. The dashed rectangle indicates the unit cell. (b) Band structure of $Si_2O$-I. The left inset shows the Brillion zone with the high symmetry points. The right inset shows the zoom-in view of the band structure corresponding to the region shaded in red. (c) Phonon spectrum of $Si_2O$-I.

**Table I.** Classification of electronic property of honeycomb materials modified by group-VI elements.

|              | O                     | S             | Se            |
|--------------|-----------------------|---------------|---------------|
| **graphene** | semiconductor         | -             | -             |
| **silicene** | semi-Dirac semimetal  | semiconductor | semiconductor |
| **germanene**| Dirac semimetal       | metal         | metal         |
| **stanene**  | Dirac semimetal       | metal         | metal         |

Given that the silicene oxide is the most promising candidate of semi-Dirac semimetals, we next focus our discussion on this material. We first study its kinetic stability. Figure 2c shows its phonon spectrum. It can be seen that over the whole BZ, no imaginary frequency (a signal of kinetic instability) was observed. We also studied the relative thermodynamic stability with other possible competing phases. For example, graphene epoxide is more stable in a (1×2) structure [36]. Our calculation shows that silicene oxide is more stable in the (1×1) structure, as shown in Figure 2a. Another two possible structures, as will be discussed below, also have higher energy than $Si_2O$-I. The chemisorption energy of $Si_2O$-I from silicene and $O_2$ gas is calculated to be -1.79 eV per formula unit of $Si_2O$, suggesting that the covalent addition of O on silicene is an exothermic process.

The enhancement of $t_2$ is correlated to the change in Si-Si bond length after oxygen adsorption. The length of the Si-Si bond underneath the O adatom in Si$_2$O-I (corresponding to $t_2$) is slightly changed from that in silicene (2.28 Å) to 2.29 Å. In contrast, the length of the Si-Si bond without O adsorption (corresponding to $t_1$) is weakened and significantly increased to 2.35 Å. This change in bond lengths effectively increases the ratio of $t_2/t_1$ and shifts the two Dirac points towards the Γ-point. Figure 2b shows the band structure of Si$_2$O-I. It is interesting to see that the two Dirac points are nearly merged to the Γ-point along the Γ-Y direction. The band dispersion is highly anisotropic. Along Γ-X, it is linear; while along Γ-Y it is parabolic, satisfying the requirement for a semi-Dirac point.

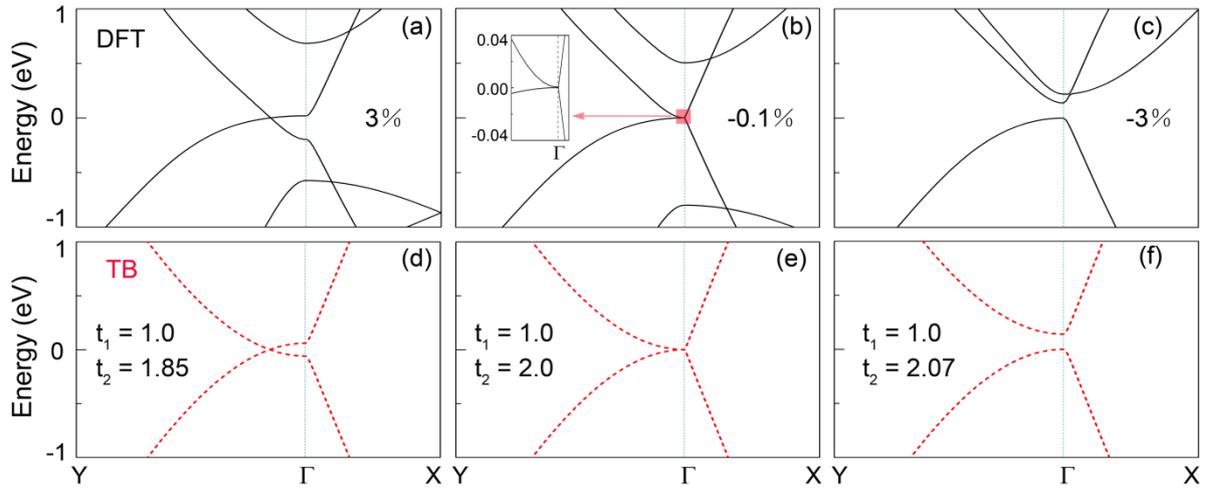

**Figure 3.** (a)-(c) Band structures of DFT of Si$_2$O-I under different uniaxial strain along the *a* direction with 3%, -0.1% and -3%, respectively. The positive (negative) value means tensile (compressive) strain. The inset in (b) is the zoom-in view of the band structure corresponding to the region shaded in red. (d-f) The TB results of the band structures corresponding to (a-c).

The remaining deviation from the semi-Dirac point (about 5 meV from our calculation) can be fine-tuned by strain applied along the armchair direction (or *a* direction in Figure 2a). Figure 3 show the band structures of Si$_2$O-I from DFT (a-c) and TB (d-f) calculations under different strains. In the TB calculations, we fix $t_1$ to 1.0 eV and alter the value of $t_2$ according to the applied strain. As DFT calculation already produces a band structure very close to a semi-Dirac point, a small compressive strain along the *a* direction would lead to a nearly perfect semi-Dirac semimetal, as shown in Figure 3b and the inset. The required strain is about -0.1%, which is much smaller than that needed in pristine graphene or silicene.

In phase Si$_2$O-I, the dispersion along Γ-X is indeed linear around Γ, and the Fermi velocities for electrons and holes are 9.6 and 9.3×10$^5$ m/s, respectively, which are even larger than that calculated in graphene (8.4×10$^5$ m/s). When the tensile strain is applied, $t_2$ will reduce and the semi-Dirac point separates into two Dirac points as shown in Figure 3a, which shows the result with 3% tensile strain corresponding to $t_2$ = 1.85 eV (Figure 3d). The location of the Dirac points on Γ-Y is given by $\pm\frac{1}{b}\arccos\left(\frac{t_2^2}{2t_1^2}-1\right)$, where $b$ is the lattice constant along the *b* direction (cf. Figure 2a), according to our TB model. Under large compressive strain, the value of $t_2$ will exceed the critical value to open a gap, as shown in Figure 3c, which is the result with -3% strain corresponding to $t_2$ = 2.07 eV (Figure 3f).

Since strain is a useful knob tuning the electronic structure, we study the critical strain under which the phase Si$_2$O-I may undergo a phase change. As shown in Figure 4, when the strain along the armchair direction is increased to about 14%, a phase change occurs. The new phase is called Si$_2$O-II with its structure shown in the inset of Figure 4. The new phase is less stable than the Si$_2$O-I phase by about 0.2 eV per unit cell. The Si$_2$O-II phase is a Dirac semimetal with its band structure shown in SI Figure S2. It is worth of mentioning that Si$_2$O-II has another related phase, named as Si$_2$O-III, as shown in SI Figure S3a. It has similar lattice constants ($a$ = 8.43 Å, $b$ = 3.95 Å) to Si$_2$O-II, but lower symmetry. This third phase is nearly degenerate with Si$_2$O-II in terms of thermodynamic stability. Its band structure is also similar to Si$_2$O-II, as shown in SI Figure S3b. Thus, among all the possible structures that we examined, Si$_2$O-I is the most stable structure.

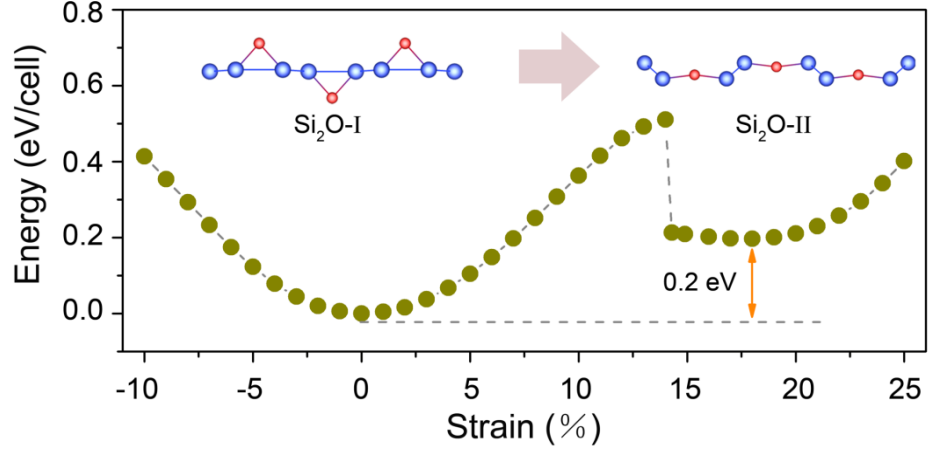

**Figure 4.** Total energy of Si$_2$O-I as a function of uniaxial strain along the *a* direction. As the compressive strain is increased to about 14%, a transition to the phase Si$_2$O-II occurs. The inset shows the structural change from Si$_2$O-I to Si$_2$O-II.

To gain insights on this peculiar band structure of Si$_2$O-I, we examined the wavefunctions of the two states at the semi-Dirac point. As seen in Figure 2a, the O adatom and two underlying Si atoms form a triangular (or epoxide) structure. In such a structure, it is known that the *p*-orbital frustration can give rise to rather uncommon bonding features [37, 38]. The bonding and anti-bonding states associated with this structure are schematically shown in Figures 5a and 5b (on the right side of the arrows), which are significantly different from that in pristine silicene, i.e., standard π-bond between two *p* orbitals (on the left side of the arrows in Figures 5a and 5b). The key feature is that the two Si $p_z$ states are significantly tilted towards each other so that the π-bonding as in pristine silicene becomes more like σ-bonding. Such a change strengthens the interaction between the two Si atoms and thus increases the hopping energy $t_2$. Correspondingly, the Dirac semimetal transits to a semi-Dirac semimetal. To better see the effect of *p*-orbital frustration, we compare the band structure with that of Si$_2$O-II, where the triangular epoxide geometry is destroyed and the *p*-orbital frustration is absent. As a result, the Si $p_z$ orbitals in Si$_2$O-II are similar to those in pristine silicene. The band structure becomes also similar to silicene, i.e., with two Dirac points located in the Γ-Y direction far away from the Γ point (See SI Figure S2b).

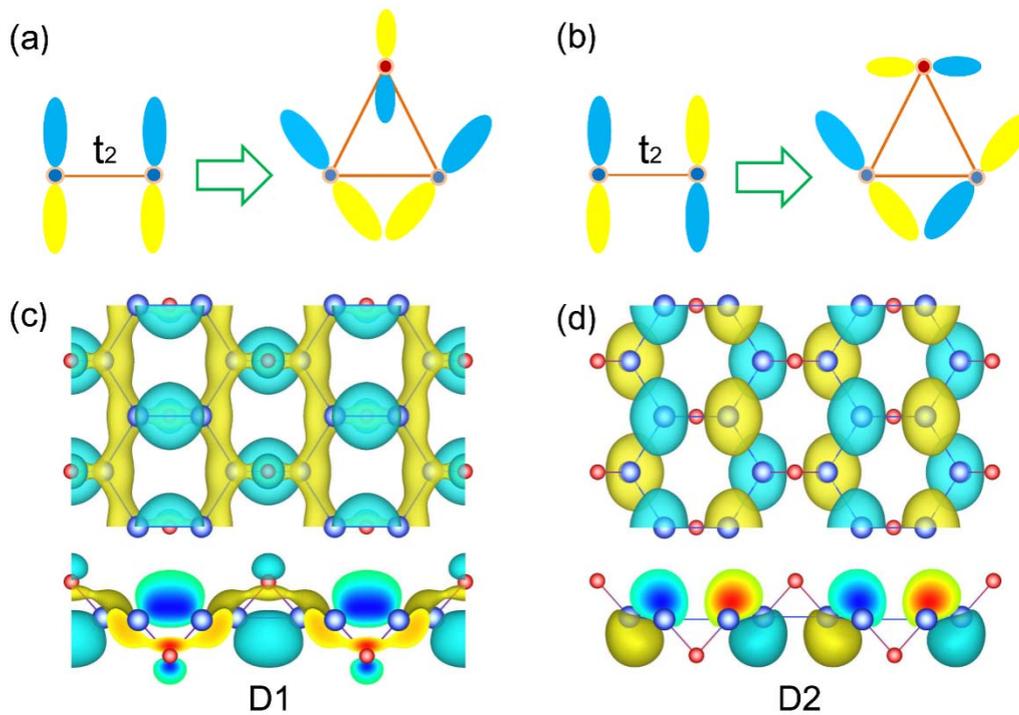

**Figure 5.** (a)-(b) Schematic illustration of bonding and anti-bonding states when changing from silicene to Si$_2$O-I. The red and blue dots represent O and Si atoms, respectively. (c)-(d) Wavefunctions of the two states D1 and D2 of Si$_2$O-I, as indicated in the inset of Figure 2b. Top views are at the top and side views are at the bottom. Blue and yellow colors represent the positive and negative parts of the wavefunctions, respectively.

The size of the adatom has significant effect on the electronic structure. As can be seen in Table I and SI Figure S1, changing the adatom from O to S on silicene will open a large band gap meaning that S has more pronounced effect on increasing the hopping energy $t_2$. This can be understood by the bonding mechanism discussed above. One can consider that the tilting of the $p_z$ orbital of Si as shown in Figures 5a-5b is a result of the presence of O/S/Se $p$ orbitals. The larger group-VI atoms would lead to more significant tilting of the Si $p_z$ orbitals towards the σ-like bonding due to a steric effect.

As S/Se is adsorbed on silicene, the honeycomb lattice forms buckling, which is different from the case of Si$_2$O-I. As a result, two $p_x$/$p_y$ states of Si enter the band gap. In this case, the band edge states are not dominated by $p_z$ orbitals anymore. The same situation occurs in the case of covalent addition of group-VI atoms to germanene and stanene. Even though the covalent addition of S/Se pushes the Dirac points closer to the Γ point, other bands induced by $p_x$/$p_y$ orbitals cross the Fermi level and lead to a metallic phase. Therefore, among the materials formed by covalent addition of group-VI atoms to the honeycomb lattice, Si$_2$O-I is the most appropriate candidate for the semi-Dirac semimetal.

In summary, using first-principles computation we searched for semi-Dirac semimetals based on graphene-like materials through a synergic application of covalent addition and strain engineering. The triangular epoxide bond with $p$-orbital frustration was found to significantly modify the hopping energies of adjacent bonds in the underlying honeycomb lattice. As a result, the two Dirac points were found to shift towards the Γ point. Then, applied strain along the armchair direction can fine-tune the band structure to obtain the semi-Dirac semimetals. Our computational search reveals that a silicene oxide is a promising candidate to realize the semi-Dirac semimetal.

Our study thus provides a condensed-matter platform for studying the new physics when massless and massive electrons coexist at the same point in the momentum space. The approach demonstrated in this work is also expected to inspire the research for discovering new semi-Dirac semimetals and beyond.

## Acknowledgements

The authors thank H. Wang for his help on the analysis of some of the calculation results. This work was supported by the National Natural Science Foundation of China (Nos. 51376005 and 11474243). The work at RPI was supported by the U.S. Department of Energy (DOE) under Grant No. DE-SC0002623. The supercomputer time was provided by the National Energy Research Scientific Computing Center (NERSC) under DOE Contract No. DE-AC02-05CH11231 and the Center for Computational Innovations (CCI) at RPI.